\documentclass[prc,preprint,showpacs,preprintnumbers,amsmath,amssymb]{revtex4}

\usepackage[mathscr]{eucal}
\usepackage{graphicx}
\def\slr#1{\setbox0=\hbox{$#1$}           % set a box for #1
   \dimen0=\wd0                                 % and get its size
   \setbox1=\hbox{/} \dimen1=\wd1               % get size of /
   \ifdim\dimen0>\dimen1                        % #1 is bigger
      \rlap{\hbox to \dimen0{\hfil/\hfil}}      % so center / in box
      #1                                        % and print #1
   \else                                        % / is bigger
      \rlap{\hbox to \dimen1{\hfil$#1$\hfil}}   % so center #1
      /                                         % and print /
   \fi}

\def\ksq{k^2}

\def\mytint#1{\!\int\!\!\frac{d^3\!{#1}}{(2\pi)^3}\,}
\def\gev#1{ GeV${}^{#1}$}
\def\be{\begin{eqnarray}}
\def\ee{\end{eqnarray}}

\renewcommand{\theequation}%
    {\arabic{section}.\arabic{equation}}
\makeatletter \@addtoreset{equation}{section} \makeatother

\begin{document}

% Use the \preprint command to place your local institutional report
% number in the upper righthand corner of the title page in preprint mode.
% Multiple \preprint commands are allowed.
% Use the 'preprintnumbers' class option to override journal defaults
% to display numbers if necessary
\preprint{BCCNT: 02/101/316}

%Title of paper
\title{Calculation of the Pseudoscalar-Isoscalar Hadronic Current Correlation Functions of
 the Quark-Gluon Plasma}

\author{Bing He}
\author{Hu Li}
\author{C. M. Shakin}
 \email[email:]{casbc@cunyvm.cuny.edu}
\author{Qing Sun}

\affiliation{%
Department of Physics and Center for Nuclear Theory\\
Brooklyn College of the City University of New York\\
Brooklyn, New York 11210
}%

\date{October, 2002}

\begin{abstract}
We report the results of calculations of pseudoscalar-isoscalar
hadronic current correlators using the Nambu--Jona-Lasinio model
and the real-time finite-temperature formalism. (This work
represents a continuation of results reported previously for other
current correlators.) Results are presented for the temperatures
range 1.2 $\leq T/T_c\leq$ 6.0, where $T_c$ is the temperature of
the confinement-deconfinement transition, which we take to be
$T_c=170$ MeV. Some resonant features are seen in our
calculations. In order to understand the origin of these
resonances, we have performed relativistic random phase
approximation (RPA) calculations of the temperature-dependent
spectrum of the $\eta$ mesons for $T<T_c$. For the RPA
calculations, use is made of a simple model in which we introduce
temperature-dependent constituent quark masses calculated in a
mean-field approximation and a temperature-dependent confining
interaction whose form is motivated by recent studies made using
lattice simulations of QCD with dynamical quarks. We also
introduce temperature-dependent coupling constants in our
generalized NJL model. Our motivation in the latter case is the
simulation of the approach to a weakly interacting system at high
temperatures and the avoidance of ``$\eta$ condensates" which
would indicate instability of the ground state of the model. We
present some evidence that supports our use of
temperature-dependent coupling constants for the NJL model. We
suggest that our results may be of interest to researchers who use
lattice simulations of QCD to obtain temperature dependent
spectral functions for various hadronic current correlation
functions.
\end{abstract}

% insert suggested PACS numbers in braces on next line
% insert suggested keywords - APS authors don't need to do this
%\keywords{}

%\maketitle must follow title, authors, abstract, \pacs, and \keywords
\maketitle

% body of paper here - Use proper section commands
% References should be done using the \cite, \ref, and \label commands
\section{INTRODUCTION}

Recently we have seen a good deal of interest in the calculation
of hadronic current correlation functions at finite temperature
using lattice simulations of QCD [1-5]. The procedure involves the
calculation of the correlation function in Euclidean space and the
application of a generalized Laplace transform to obtain the
spectral functions. Various temperature-dependent resonant
structures are found in the spectral functions. It is of interest
to obtain further understanding of the nature of these resonances
using chiral Lagrangian models, such as that of the
Nambu--Jona-Lasinio (NJL) model [6], that are successful in
reproducing some of the low-energy properties of QCD. We have made
some calculations of spectral functions making use of a
generalized Nambu--Jona-Lasinio model [7]. In that work we studied
correlators for currents with the quantum numbers of the $a_0$,
$f_0$, $\pi$, and $\rho$ mesons. In the present work we extend our
study to the pseudoscalar-isoscalar correlators. In this case the
correlators are calculated in terms of excitations that have the
quantum numbers of the $\eta$ mesons. One of our goals in the
present study is to relate the pseudoscalar-isoscalar spectral
functions to the properties of the $\eta$ mesons at finite
temperature. The temperature dependence of the spectrum of the eta
mesons is obtained using our generalized NJL model which includes
a covariant model of confinement [8-13]. At zero temperature our
model provides an excellent fit to the properties of the
$\eta$(547), $\eta^\prime$(958) mesons and their radial
excitations [12]. (The model gives quite satisfactory results for
the decay constants and mixing angles of the $\eta$ and
$\eta^\prime$ mesons.) In order to obtain the spectrum of the eta
mesons at finite temperature, we make some assumptions which lead
to relatively simple calculations. For example, we use
temperature-dependent mass values for the constituent quarks
obtained in a mean-field analysis and also introduce a
temperature-dependent confining interaction that is motivated by
some recent results obtained in lattice simulations of QCD with
dynamical quarks. In addition, we have introduced a temperature
dependence of the coupling constants of the NJL model. Our
original motivation for introducing that temperature dependence
was to model the various physical mechanisms that work against the
development of pion condensation [14]. However, in the present
work we argue that, if the NJL model is to be used at high
temperatures, such temperature dependence is needed to avoid
results that are inconsistent with what is known about QCD
thermodynamics. For our studies we have introduced coupling
constants $G(T)=G(0)[1-0.17\,T/T_c]$, where $T_c$ is the
temperature of the confinement-deconfinement transition, which we
take to be $T_c=170$ MeV. At that temperature the coupling
constants are reduced by 17\%. The coupling constants are equal to
zero beyond $T=5.88\,T_c$. Thus, we see that in this model the
short-range interaction is present in the range 0 $\leq T \leq$
5.88\,$T_c$, while the confining interaction is taken to vanish
for $T\geq1.2\,T_c$. That feature of the model is also consistent
with what is known concerning QCD thermodynamics.

Ideally, it would be preferable if we could calculate the
temperature-dependent spectrum of the $\eta$ mesons, including the
radial excitations, in the imaginary-time Matsubara formalism or
in the real-time finite-temperature formalism [15, 16]. That is a
formidable task which is beyond the scope of the present work, but
might be considered at some future time. (On the whole, there are
some unresolved questions concerning the use of the
field-theoretic finite-temperature theory in the confined phase of
QCD which are not present when considering the deconfined phase.)
In those cases in which our results may be compared to those of
the Matsubara formalism we find general agreement [17]. However,
we need to assume for the present work that, with the use of
temperature-dependent constituent quark mass values, a
temperature-dependent confining interaction and
temperature-dependent coupling constants, we have introduced the
most important features that influence the $\eta$ spectrum for
$T<T_c$.

The organization of our work is as follows. In Section II we
discuss the calculation of polarization functions of the NJL model
at finite temperature. In Section III we describe the calculation
of the pseudoscalar-isoscalar hadronic current correlation
functions and present various results of our numerical
computations. Since the use of temperature-dependent coupling
constants is an unusual feature of our model, we provide some
justification for the use of such coupling constants in Section
IV. In Section V we introduce the Lagrangian of our model that is
used for our RPA calculations. We make reference to the RPA
equations that were used to calculate the properties of the $\eta$
mesons at $T=0$ in earlier work and describe the motivation for
the introduction of a temperature-dependent confining interaction.
We present some results of our numerical calculations and relate
the results of our RPA calculations for $T<T_c$ to the results
obtained for the correlation functions for $T>T_c$. Finally,
Section VI contains some further discussion and conclusions.

\section{polarization functions at finite temperature}

In an earlier work we carried out a Euclidean-space calculation of
the up, down, and strange constituent quark masses taking into
account the 't Hooft interaction and our confining interaction
[18]. The 't Hooft interaction plays only a minor role, but does
provide coupling of the equations for the various constituent
masses. If we neglect the confining interaction and the 't Hooft
interaction in the mean field calculation of the constituent
masses, we can compensate for their absence by making a modest
change in the value of $G_S$, the coupling constant of the NJL
model. For the calculations of this work we calculate the quark
masses using the formalism presented in the Klevansky review [19].
(Note that our value of $G_S$ is twice the value of $G$ used in
that review.) The relevant equation is Eq. (5.38) of Ref. [19].
Here, we put $\mu=0$ and write \be
m(T)=m^0+4GN_c\frac{m(T)}{\pi^2}\int_0^\Lambda
dp\frac{p^2}{E_p}\tanh(\frac 1 2\beta E_p)\,,\ee where
$\Lambda=0.631$ GeV is a cutoff for the momentum integral,
$\beta=1/T$ and $E_p=[\vec p^2+m^2(T)]^{1/2}$. In our calculations
we replace $G$ by $G_S(T)/2$ and solve the equation \be
m(T)=m^0+2G_S(T)N_c\frac{m(T)}{\pi^2}\int_0^\Lambda
dp\frac{p^2}{E_p}\tanh(\frac 1 2\beta E_p)\,,\ee with
$G_S(T)=11.38[1-0.17\,T/T_c]$ GeV, $m_u^0=0.0055$ GeV and
$m_s^0=0.130$ GeV. Thus, we see that $G_S(T)$ is reduced from the
value $G_S(0)$ by 17\% when $T=T_c$. The results obtained in this
manner for $m_u(T)$ and $m_s(T)$ are shown in Fig. 1. Here, the
temperature dependence we have introduced for $G_S(T)$ serves to
provide a somewhat more rapid restoration of chiral symmetry than
that which is found for a constant value of $G_S$. That feature
and the temperature dependence of the confining potential leads to
the deconfinement of the light mesons considered here at $T\geq
T_c$.

For the calculation of polarization functions for $T>1.2\,T_c$ we
may neglect the confining interaction. However, we include
temperature-dependent quark mass values in our calculations. The
basic polarization functions that are calculated in the NJL model
are shown in Fig. 2. We will consider calculations of such
functions in the frame where $\vec P=0$. In our earlier work,
calculations were made after a confinement vertex was included.
That vertex is represented by the filled triangular region in Fig.
2. However, we here consider calculations for $T\geq 1.2T_c$ where
confinement may be neglected.

\begin{figure}
 \includegraphics[bb=0 0 300 200, angle=0, scale=1.2]{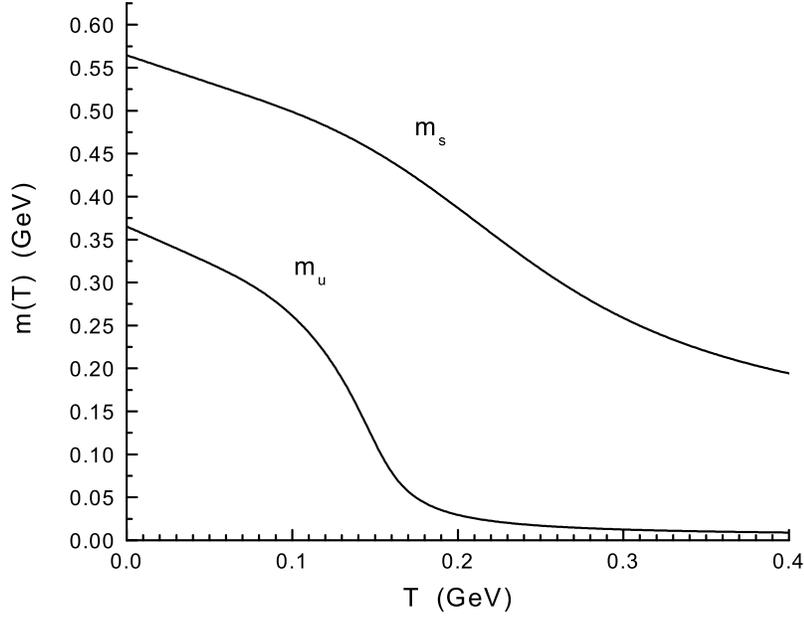}%
 \caption{Temperature dependent constituent mass values, $m_u(T)$ and $m_s(T)$,
 calculated using Eq. (2.2) are shown. Here $m_u^0=0.0055$ GeV,
 $m_s^0=0.130$ GeV, and $G(T)=5.691[1-0.17(T/T_c)]$, if we use
 Klevansky's notation [19].}
 \end{figure}

\begin{figure}
 \includegraphics[bb=0 0 600 350, angle=0, scale=0.4]{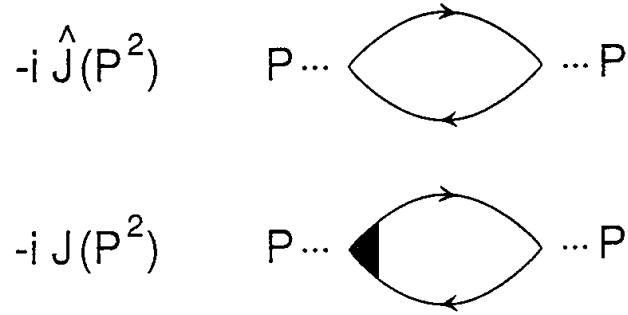}%
 \caption{The upper figure represents the basic polarization diagram of the
 NJL model in which the lines represent a constituent quark and a constituent
 antiquark. The lower figure shows a confinement vertex [filled triangular
 region] used in our earlier work. For the present work we neglect confinement
 for $T\geq1.2T_c$, with $T_c=170$ MeV.}
 \end{figure}

The procedure we adopt is based upon the real-time
finite-temperature formalism, in which the imaginary part of the
polarization function may be calculated. Then, the real part of
the function is obtained using a dispersion relation. The result
we need for this work has been already given in the work of Kobes
and Semenoff [20]. (In Ref. [20] the quark momentum in Fig. 2 is
$k$ and the antiquark momentum is $k-P$. We will adopt that
notation in this section for ease of reference to the results
presented in Ref. [20].) With reference to Eq. (5.4) of Ref. [20],
we write the imaginary part of the scalar polarization function as
\be \mbox{Im}\,J^S(P^2,
T)=\frac12(2N_c)\beta_S\,\epsilon(p^0)\mytint ke^{-\vec
k\,{}^2/\alpha^2}\left(\frac{2\pi}{2E_1(k)2E_2(k)}\right)\\\nonumber
\{(1-n_1(k)-n_2(k))
\delta(p^0-E_1(k)-E_2(k))\\\nonumber-(n_1(k)-n_2(k))
\delta(p^0+E_1(k)-E_2(k))\\\nonumber-(n_2(k)-n_1(k))
\delta(p^0-E_1(k)+E_2(k))\\\nonumber-(1-n_1(k)-n_2(k))
\delta(p^0+E_1(k)+E_2(k))\}\,.\ee Here, $E_1(k)=[\vec
k\,{}^2+m_1^2(T)]^{1/2}$. Relative to Eq. (5.4) of Ref. [20], we
have changed the sign, removed a factor of $g^2$ and have included
a statistical factor of $2N_c$, where the factor of 2 arises from
the flavor trace. In addition, we have included a Gaussian
regulator, $\exp[-\vec k\,{}^2/\alpha^2]$, with $\alpha=0.605$
GeV, which is the same as that used in most of our applications of
the NJL model in the calculation of meson properties [8-15]. We
also note that \be n_1(k)=\frac1{e^{\,\beta E_1(k)}+1}\,,\ee and
\be n_2(k)=\frac1{e^{\,\beta E_2(k)}+1}\,.\ee For the calculation
of the imaginary part of the polarization function, we may put
$\ksq=m_1^2(T)$ and $(k-P)^2=m_2^2(T)$, since in that calculation
the quark and antiquark are on-mass-shell. We will first remark
upon the calculation of scalar correlators. In that case, the
factor $\beta_S$ in Eq. (2.3) arises from a trace involving Dirac
matrices, such that
\be \beta_S&=&-\mbox{Tr}[(\slr k+m_1)(\slr k-\slr P+m_2)]\\
&=&2P^2-2(m_1+m_2)^2\,,\ee where $m_1$ and $m_2$ depend upon
temperature. In the frame where $\vec P=0$, and in the case
$m_1=m_2$, we have $\beta_S=2P_0^2(1-{4m^2}/{P_0^2})$. For the
scalar case, with $m_1=m_2$, we find \be \mbox{Im}\,J^S(P^2,
T)=\frac{N_cP_0^2}{4\pi}\left(1-\frac{4m^2}{P_0^2}\right)^{3/2}
e^{-\vec k\,{}^2/\alpha^2}[1-2n_1(k)]\,,\ee where \be \vec
k\,{}^2=\frac{P_0^2}4-m^2(T)\,.\ee

We may evaluate Eq. (2.8) for $m(T)=m_u(T)=m_d(T)$ and define
$\mbox{Im}\,J_u^S(P^2, T)$. Then we put $m(T)=m_s(T)$, we define
$\mbox{Im}\,J_s^S(P^2, T)$. These two functions are needed for a
calculation of the scalar-isoscalar correlator. (Note that the
factor of 2 arising from the flavor trace should be moved when we
define the polarization function for a specific quark flavor.) The
real parts of the functions $J_u^S(P^2, T)$ and $J_s^S(P^2, T)$
may be obtained using a dispersion relation, as noted earlier.

For pseudoscalar mesons, we replace $\beta_S$ by
\be \beta_P&=&-\mbox{Tr}[i\gamma_5(\slr k+m_1)i\gamma_5(\slr k-\slr P+m_2)]\\
&=&2P^2-2(m_1-m_2)^2\,,\ee which for $m_1=m_2$ is $\beta_P=2P_0^2$
in the frame where $\vec P=0$. We find, for the $\pi$ mesons, \be
\mbox{Im}\,J^P(P^2,T)=\frac{N_cP_0^2}{4\pi}\left(1-\frac{4m(T)^2}{P_0^2}\right)^{1/2}
e^{-\vec k\,{}^2/\alpha^2}[1-2n_1(k)]\,,\ee where $ \vec
k\,{}^2={P_0^2}/4-m_u^2(T)$, as above. Thus, we see that, relative
to the scalar case, the phase space factor has an exponent of 1/2
corresponding to a \textit{s}-wave amplitude, rather than the
\textit{p}-wave amplitude of scalar mesons. For the scalars, the
exponent of the phase-space factor is 3/2, as seen in Eq. (2.8).

For a study of vector mesons we consider \be
\beta_{\mu\nu}^V=\mbox{Tr}[\gamma_\mu(\slr k+m_1)\gamma_\nu(\slr
k-\slr P+m_2)]\,,\ee and calculate \be
g^{\mu\nu}\beta_{\mu\nu}^V=4[P^2-m_1^2-m_2^2+4m_1m_2]\,,\ee which,
in the equal-mass case, is equal to $4P_0^2+8m^2(T)$, when
$m_1=m_2$ and $\vec P=0$. Note that for the elevated temperatures
considered in this work $m_u(T)=m_d(T)$ is quite small, so that
$4P_0^2+8m_u^2(T)$ can be approximated by $4P_0^2$ when we
consider the $\rho$ meson. The generalization of these results for
the study of the pseudoscalar-isoscalar correlators will be taken
up in the next section.

\section{calculation of hadronic current correlation functions}

In this section we consider the calculation of
temperature-dependent hadronic current correlation functions. The
general form of the correlator is a transform of a time-ordered
product of currents, \be C(P^2, T)=i\int d^4xe^{\,ip\cdot
x}<<\mbox T(j(x)j(0)>>\,,\ee where the double bracket is a
reminder that we are considering the finite temperature case.

For the study of pseudoscalar states, we may consider currents of
the form $j_{P,\,i}(x)=\bar q(x)i\gamma_5\lambda^iq(x)$, where, in
the case of the $\pi$ mesons, $i=1,2,$ and 3. For the study of
pseudoscalar-isoscalar mesons, we again introduce
$j_{P,\,i}(x)=\bar q(x)\lambda^iq(x)$, but here $i=0$ for the
flavor-singlet current and $i=8$ for the flavor-octet current.

In the case of the $\pi$ mesons, the correlator may be expressed
in terms of the basic vacuum polarization function of the NJL
model, $J_P(P^2, T)$ [19, 21, 22]. Thus, \be C_\pi(P^2,T)=J_P(P^2,
T)\frac1{1-G_\pi(T)J_P(P^2, T)}\,,\ee where $G_\pi(T)$ is the
coupling constant appropriate for our study of the $\pi$ mesons.
We have found $G_\pi(0)=13.49$\gev{-2} by fitting the pion mass in
a calculation made at $T=0$, with $m_u=m_d=0.364$ GeV [14].

The calculation of the correlator for pseudoscalar-isoscalar
states is more complex, since there are both flavor-singlet and
flavor-octet states to consider. We may define polarization
functions for $u$, $d$ and $s$ quarks: $J_u(P^2, T)$, $J_d(P^2,
T)$ and $J_s(P^2, T)$. (We recall that the factor of 2 arising
from the flavor trace is not included when these functions are
calculated.) In terms of these polarization functions we may then
define \be J_{00}(P^2, T)=\frac23[J_u(P^2, T)+J_d(P^2, T)+J_s(P^2,
T)]\,,\ee \be J_{08}(P^2, T)=\frac{\sqrt2}3[J_u(P^2, T)+J_d(P^2,
T)-2J_s(P^2, T)]\,,\ee and \be J_{88}(P^2, T)=\frac13[J_u(P^2,
T)+J_d(P^2, T)+4J_s(P^2, T)]\,.\ee We also introduce the matrices
\be J(P^2, T)=\left[\begin{array}{cc}J_{00}(P^2, T)&J_{08}(P^2,
T)\\J_{80}(P^2, T)&J_{88}(P^2, T)\end{array}\right]\,,\ee \be
G(T)=\left[\begin{array}{cc}G_{00}(T)&G_{08}(T)\\G_{80}(T)&G_{88}
(T)\end{array}\right]\,,\ee and \be C(P^2,
T)=\left[\begin{array}{cc}C_{00}(P^2, T)&C_{08}(P^2,
T)\\C_{80}(P^2, T)&C_{88}(P^2, T)\end{array}\right]\,.\ee We then
write the matrix relation \be C(P^2, T)=J(P^2, T)[1-G(T)J(P^2,
T)]^{-1}\,.\ee

For some purposes it may be useful to also define a \emph{t}
matrix \be t(P^2, T)=[1-G(T)J(P^2, T)]^{-1}G(T)\,,\ee where
$t(P^2, T)$ has the structure shown in Eqs. (3.6)-(3.8). The same
resonant structures are seen in both $C(P^2, T)$ and $t(P^2, T)$.

Some of our results for the imaginary parts of the
pseudoscalar-isoscalar correlators $C_{00}(P^2)$, $C_{88}(P^2)$
and $C_{08}(P^2)$ are shown in Figs. 3, 4 and 5, respectively. In
these figures the values are $T/T_c=1.2$ [solid line], $T/T_c=1.6$
[dashed line], $T/T_c=2.0$ [dotted line], $T/T_c=4.0$
[dashed-dotted line] and $T/T_c=6.0$ [dashed-(double)dotted line].
There is a large peak seen in Figs. 3-5 at about 775 MeV. It is
worth noting that the state that evolves from the $\eta^\prime
(958)$ with increasing temperature has a mass of about 750 MeV for
$T\simeq T_c$. However, an analysis of the mixing angle for the
state at 775 MeV shows that it is mainly an $s\bar s$ state.
Further work is needed to understand the relation between the
bound-state spectrum for $T<T_c$ and the resonant structures seen
for $T>T_c$. We discuss the temperature dependence of the $\eta$
spectrum in the Section V.

\begin{figure}
\includegraphics[bb=0 0 300 200, angle=0, scale=1.2]{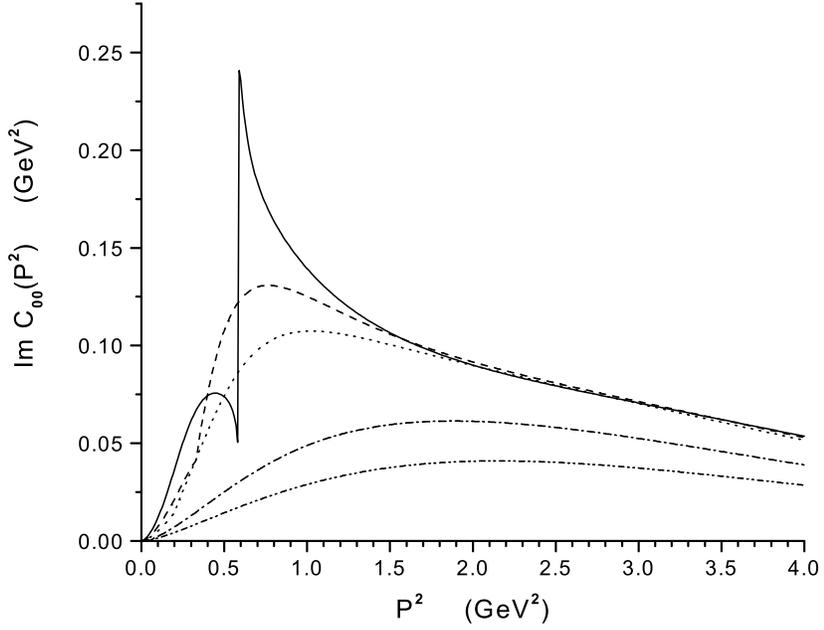}%
\caption{The imaginary part of the pseudoscalar-isoscalar
correlator $C_{00}(P^2)$ is shown. Here, $T/T_c=1.2$ [solid line],
1.6 [dashed line], 2.0 [dotted line], 4.0 [dashed-dotted line] and
6.0 [dashed-(double)dotted line]. In this work we use
$G_{00}=8.09$\gev{-2}, $G_{88}=13.02$\gev{-2} and
$G_{08}=-0.4953$\gev{-2}.}
\end{figure}

\begin{figure}
\includegraphics[bb=0 0 300 200, angle=0, scale=1.2]{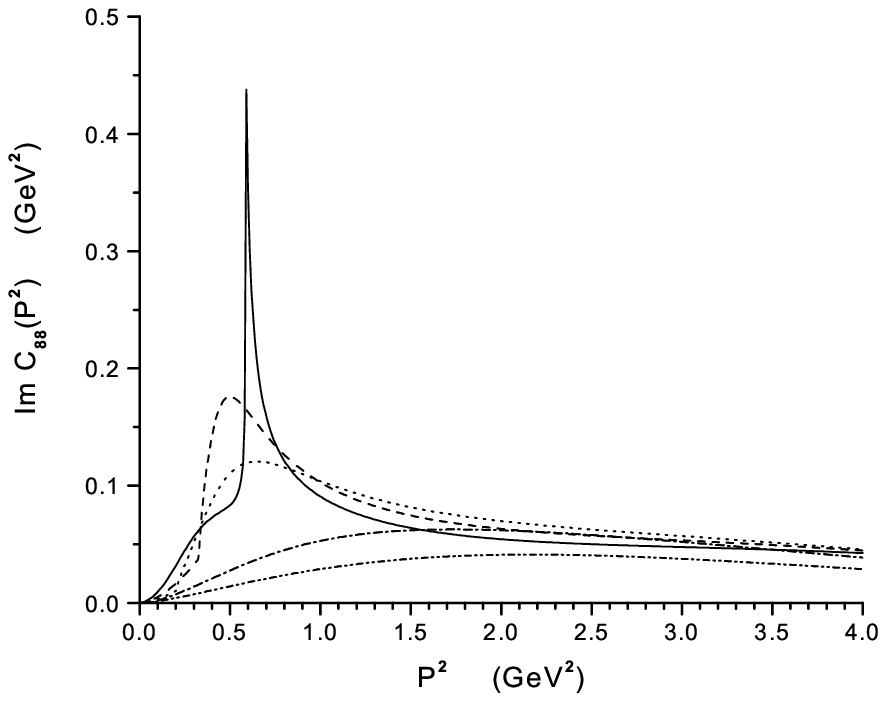}%
\caption{The imaginary part of the correlator $C_{88}(P^2)$ is
shown. [See caption to Fig. 3.]}
\end{figure}

\begin{figure}
\includegraphics[bb=0 0 300 200, angle=0, scale=1.2]{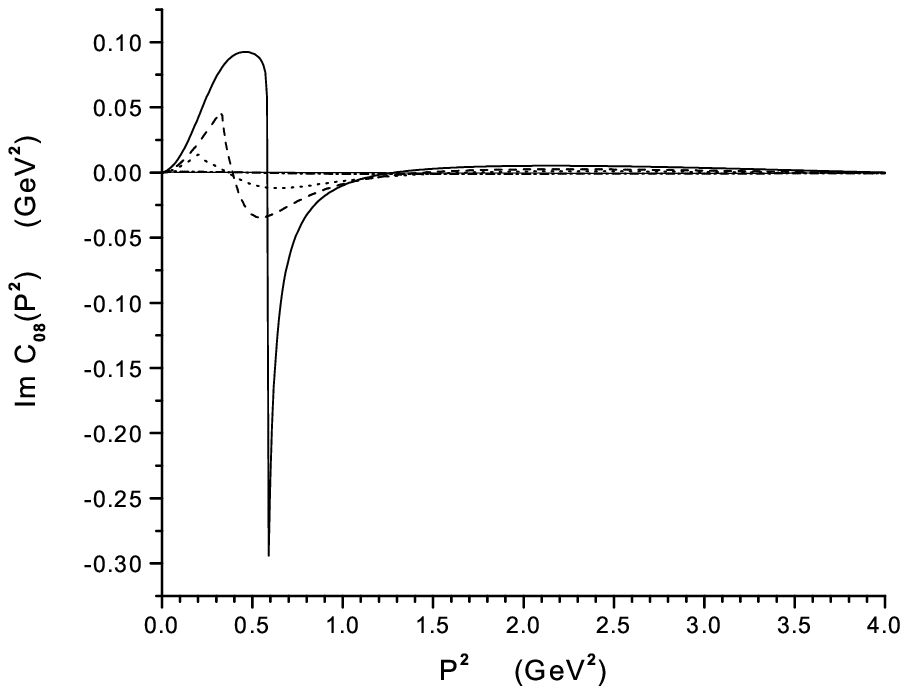}%
\caption{The imaginary part of the correlator $C_{08}(P^2)$ is
shown. [See caption to Fig. 3.]}
\end{figure}

\section{temperature-dependent coupling constants of the NJL model}

Since the introduction of temperature-dependent coupling constants
for the NJL model is a novel feature of our work, we provide
arguments in this section to justify their introduction. We make
reference to Fig. 1.3 of Ref. [16]. That figure shows the behavior
of the ratio $\epsilon/T^4$ and $3P/T^4$ for the pure gauge sector
of QCD. Here $\epsilon$ is the energy density and $P$ is the
pressure. Ideal gas behavior implies $\epsilon=3P$. The values of
$\epsilon/T^4$ and $3P/T^4$ are compared to the value
$\epsilon_{SB}/T^4=8\pi^2/15$ for an ideal gluon gas. It may be
seen from the figure that at $T=3T_c$ there are still significant
differences from the ideal gluon gas result. Deviations from ideal
gas behavior become progressively smaller with increasing $T/T_c$
and could be considered to be relatively unimportant for
$T/T_c>5$.

The use of our energy-dependent coupling constants is meant to be
consistent with the approach to asymptotic freedom at high
temperature. In order to understand this feature in our model, we
can calculate the correlator with \emph{constant} values of
$G_{00}$, $G_{88}$ and $G_{08}$ and with
$G_{00}(T)=G_{00}[1-0.17\,T/T_c]$, etc. (In this work we use
$G_{00}=8.09$\gev{-2}, $G_{88}=13.02$\gev{-2} and
$G_{08}=-0.4953$\gev{-2}.)

We now consider the values of $\mbox{Im}\,C_{88}(P^2)$ for
$T/T_c=4.0$. In Fig. 6 we show the values of
$\mbox{Im}\,C_{88}(P^2)$ calculated in our model with
temperature-dependent coupling constants as a dashed line. The
dotted line shows the values of the correlator for
$G_{00}=G_{88}=G_{08}=0$, while the solid line shows the values
when the coupling constants are kept at their values at $T=0$. We
see that we have some resonant behavior in the case the constants
are temperature independent.

In Fig. 7 we show similar results for $T/T_c=5.88$. Here the
temperature-dependent coupling constants are equal to zero, so
that the lines corresponding to the dashed and dotted lines of
Fig. 6 coincide. The solid line again shows some resonant behavior
at a value of $T/T_c$, where we expect only very weak interactions
associated with asymptotic freedom. We conclude that the model
with constant values of the coupling constants yields unacceptable
results, while our model, which has temperature-dependent coupling
constants, behaves as one may expect, when the results of lattice
simulations of QCD thermodynamics are taken into account.

\begin{figure}
\includegraphics[bb=0 0 300 200, angle=0, scale=1.2]{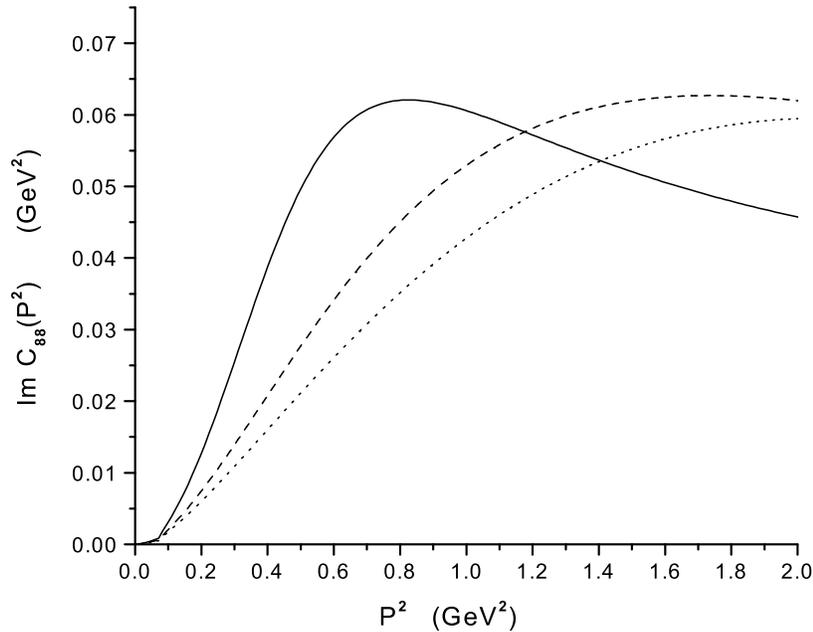}%
\caption{The imaginary part of the correlator $C_{88}(P^2)$ is
shown for $T/T_c=4.0$. The dashed line is the result for the
temperature-dependent coupling constants of our model, while the
solid line represents the results for coupling constants kept at
their $T=0$ values. [See caption to Fig. 3.] The dotted line shows
the values of the correlator when the coupling constants are set
equal to zero.}
\end{figure}

\begin{figure}
\includegraphics[bb=0 0 300 200, angle=0, scale=1.2]{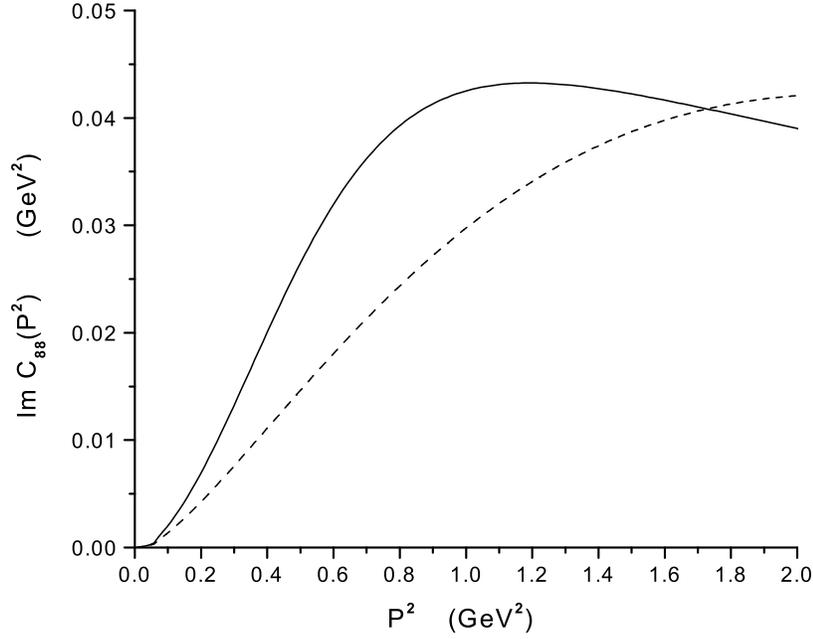}%
\caption{The imaginary part of the correlator $C_{88}(P^2)$ is
shown for $T/T_c=5.88$. [See caption to Fig. 6.] Here the dashed
and dotted lines of Fig. 6 coincide.}
\end{figure}

\section{calculation of meson properties at finite temperature in a
generalized NJL model with confinement}

It is useful to record the Lagrangian used in our calculations of
meson properties \be {\cal L}=&&\bar q(i\slr
\partial-m^0)q +\frac{G_S}{2}\sum_{i=0}^8[
(\bar q\lambda^iq)^2+(\bar qi\gamma_5 \lambda^iq)^2]\nonumber\\
&&-\frac{G_V}{2}\sum_{i=0}^8[
(\bar q\lambda^i\gamma_\mu q)^2+(\bar q\lambda^i\gamma_5 \gamma_\mu q)^2]\nonumber\\
&& +\frac{G_D}{2}\{\det[\bar q(1+\gamma_5)q]+\det[\bar
q(1-\gamma_5)q]\} \nonumber\\
&&+ {\cal L}_{conf}\,, \ee Here, $m^0$ is a current quark mass
matrix, $m^0=\mbox{diag} \,(m_u^0, \,m_d^0, \,m_s^0)$. The
$\lambda^i$ are the Gell-Mann (flavor) matrices,
$\lambda^0=\sqrt{2/3}\mathbf{\,1}$ with $\mathbf{\,1}$ being the
unit matrix. The fourth term on the right-hand side of Eq. (5.1)
is the 't Hooft interaction. Finally, ${\cal L}_{conf}$ represents
the model of confinement we have used in our work.

As noted earlier, we have recently reported results of our
calculations of the temperature dependence of the spectra of
various mesons [14]. These calculations were made using our
generalized NJL model which includes a covariant model of
confinement. We have presented results for the $\pi$, $K$, $a_0$,
$f_0$ and $K_0^*$ mesons in Ref. [14]. The equations that we solve
are of the form of relativistic random-phase-approximation
equations. The derivation of these equations for pseudoscalar
mesons is given in Ref. [13], where we discuss the equations for
pionic, kaonic and eta mesons. The equations for the eta mesons
are the most complicated, since we consider singlet-octet mixing
as well as pseudoscalar--axial-vector mixing. In that case there
are eight vertex functions to consider, $\Gamma_{P,0}^{+-}$,
$\Gamma_{A,0}^{+-}$, $\Gamma_{P,8}^{+-}$, $\Gamma_{A,8}^{+-}$,
$\Gamma_{P,0}^{-+}$, $\Gamma_{A,0}^{-+}$, $\Gamma_{P,8}^{-+}$,
$\Gamma_{A,8}^{-+}$, where $P$ refers to the $\gamma_5$ vertex and
$A$ refers to the $\gamma_0\gamma_5$ vertex, which mixes with the
$\gamma_5$ vertex. Corresponding to the eight vertex functions one
may define eight wave function amplitudes [13]. Since the RPA
equations for the study of the eta meson are quite lengthy [13],
we do not reproduce them here.

In the RPA equations we replace $m_u$ and $m_s$ by $m_u(T)$ and
$m_s(T)$ of Fig. 1 and use the temperature-dependent coupling
constants described earlier in this work. In addition to the
temperature dependence of the coupling constants and constituent
mass values, we also introduced a temperature-dependent confining
potential, whose form was motivated by recent lattice simulations
of QCD in which the temperature dependence of the confining
interaction was calculated with dynamical quarks [23]. (See Fig.
8.) In order to include such effects, we modified the form of our
confining interaction, $V^C(r)=\kappa r\exp[-\mu r]$, by replacing
$\mu$ by \be
\mu(T)=\frac{\mu_0}{\left[1-0.7\left(\displaystyle\frac{T}{T_c}\right)^2\right]}\,,\ee
with $\mu_0=0.010$ GeV. The maximum value of $V^C(r, T)$ is then
\be
V_{max}^C(T)&=&\frac\kappa{\mu(T)e}\,,\\&=&\frac{\kappa\left[1-0.7({T}/{T_c})^2\right]}
{\mu_0e}\,,\ee with $r_{max}=1/\mu(T)$. To better represent the
qualitative features of the results shown in Fig. 8, we use
$V^C(r, T)=\kappa r\exp[-\mu(T) r]$ for $r\leq r_{max}$ and
$V^C(r, T)=V_{max}^C(T)$ for $r>r_{max}$. We also note that we use
Lorentz-vector confinement and carry out all our calculations in
momentum-space. (The value of $\kappa$ used in our work is
0.055\gev2.) Values of $V^C(r, T)$ are shown is Fig. 9.

\begin{figure}
 \includegraphics[bb=0 0 250 350, angle=-90, scale=1.2]{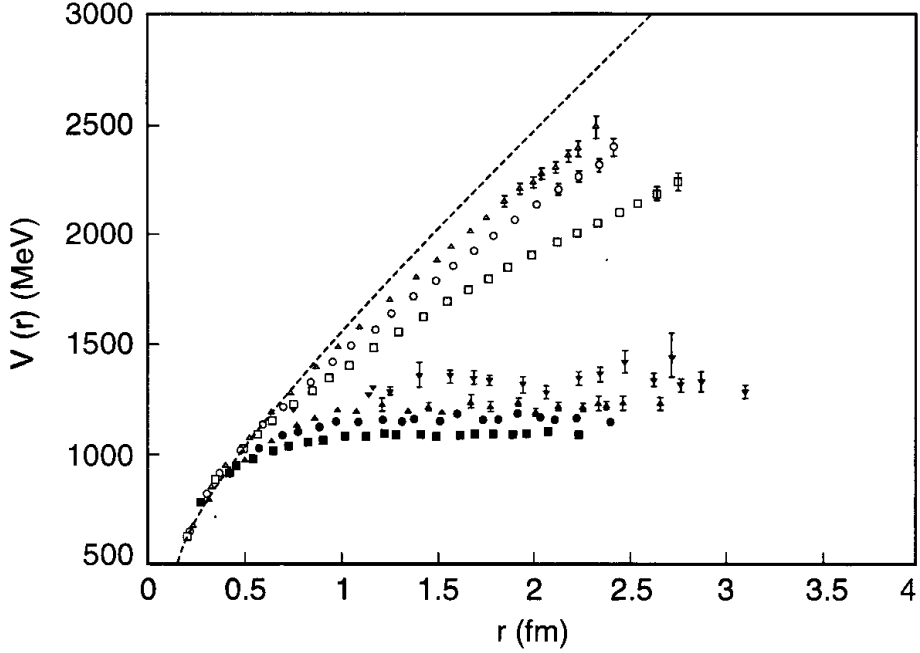}%
 \caption{A comparison of quenched (open symbols) and unquenched (filled symbols) results for
 the interquark potential at finite temperature [23]. The dotted line is the zero temperature
 quenched potential. Here, the symbols for $T=0.80T_c$ [open triangle], $T=0.88T_c$
 [open circle], and $T=0.94T_c$ [open square], represent the quenched
 results. The results with dynamical fermions are given at $T=0.68T_c$ [solid downward-pointing
 triangle], $T=0.80T_c$ [solid upward-pointing triangle], $T=0.88T_c$ [solid circle],
 and $T=0.94T_c$ [solid square].}
 \end{figure}

 \begin{figure}
 \includegraphics[bb=0 0 300 200, angle=0, scale=1.2]{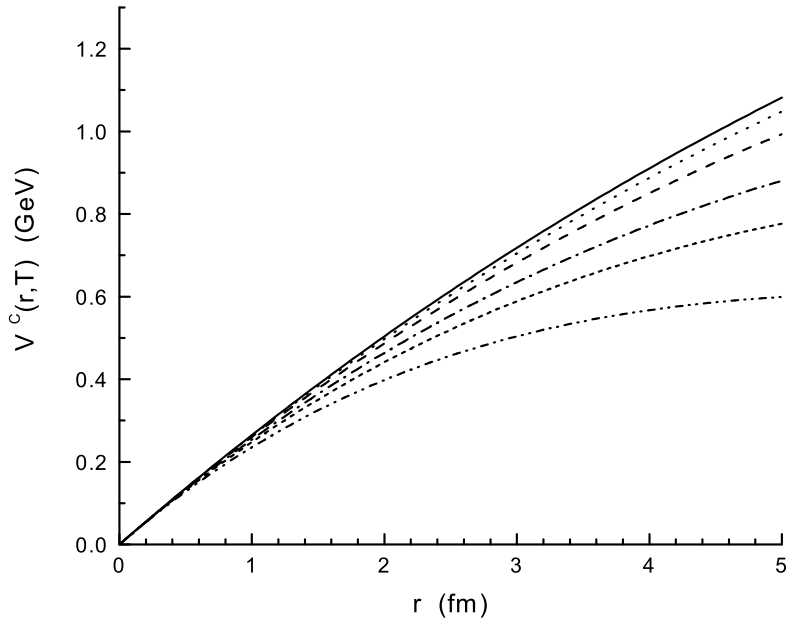}%
 \caption{The potential $V^C(r, T)$ is shown for $T/T_c=0$ [solid line],
 $T/T_c=0.4$ [dotted line], $T/T_c=0.6$ [dashed line], $T/T_c=0.8$ [dashed-dotted line],
 $T/T_c=0.9$ [short dashes], $T/T_c=1.0$ [dashed-(double) dotted line]. Here,
 $V^C(r,T)=\kappa r\exp[-\mu(T)r]$, with $\mu(T)=0.01\mbox{GeV}/[1-0.7(T/T_c)^2]$ and
 $\kappa=0.055$\gev2.}
 \end{figure}

In Fig. 10 we show the results of our calculations of the
temperature-dependent spectrum of the $\eta$ mesons. We show the
behavior of the $\eta$(547), $\eta^\prime$(958) and seven states
which represent radial excitations. The energies of the additional
states found when diagonalizing the RPA Hamiltonian are
represented by dots in the range 0 $\leq T/T_c \leq$ 0.6. We note
that the masses of the nodeless states (the $\eta$ and
$\eta^\prime$) are fairly constant over a broad range of
temperatures. That characteristic seems to be a feature of the
behavior of pseudo-Goldstone bosons at finite temperature.

\begin{figure}
 \includegraphics[bb=0 0 500 600, angle=0, scale=0.4]{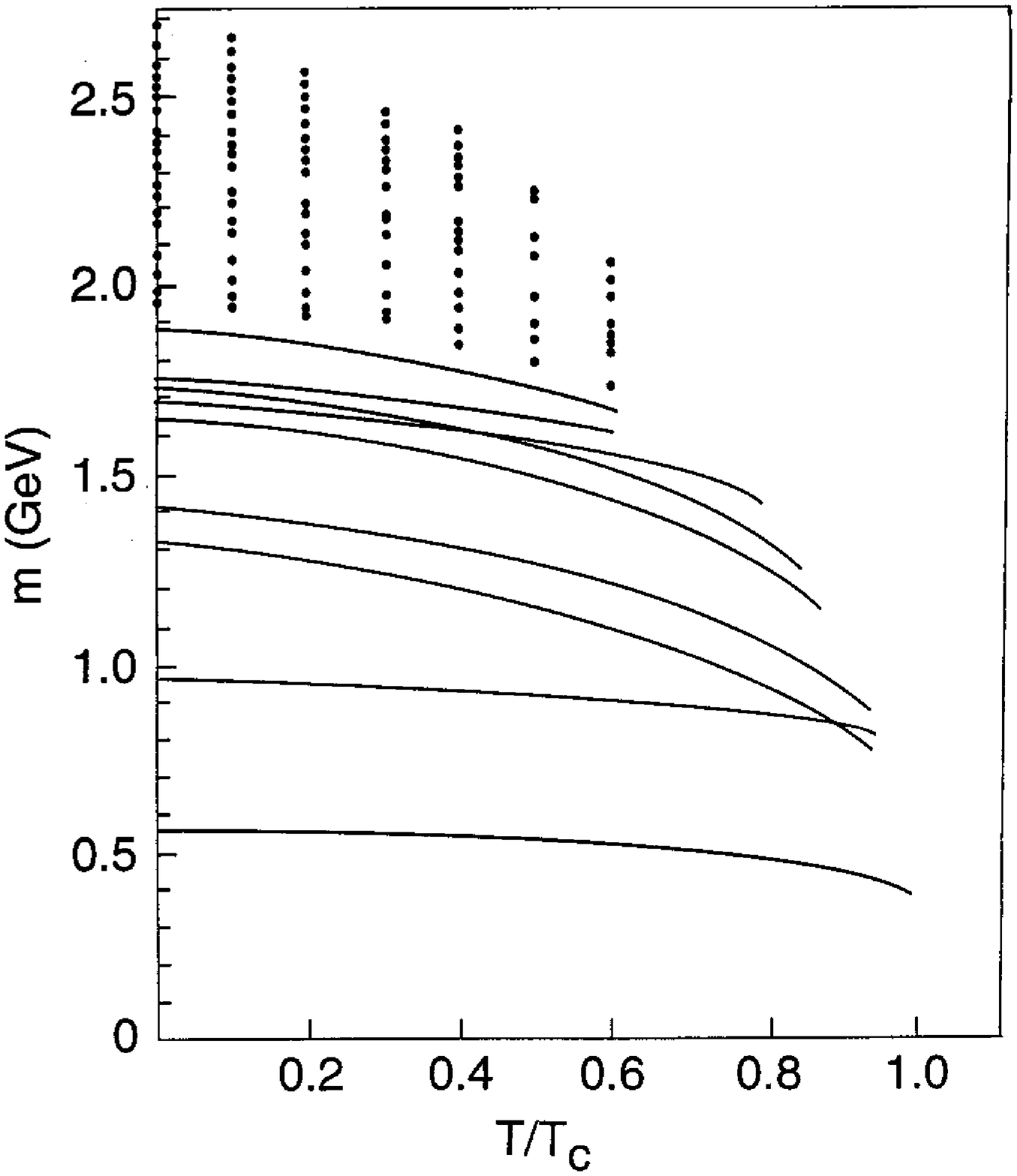}
 \caption{The temperature-dependent spectrum of the $\eta$ mesons
 is shown. For the most highly excited radial excitations we represent
 the mass values obtained by dots. There are no bound states for $T>T_c$.}
 \end{figure}

As the temperature is increased, fewer states are bound by the
confining field which decreases in magnitude with increasing
temperature. At $T=T_c$ only the state that evolves from the
$\eta$(547) is bound. That state disappears from the spectrum for
$T>T_c$. We believe that the crossing of levels seen at
$T/T_c=0.9$ is due to the rapid decrease of $m_u(T)$ relative to
$m_s(T)$ with increasing $T/T_c$. [See Fig. 1.] That feature could
lead to a (predominately) $n\bar n$ state with a node to have a
lower energy than a (predominantly) nodeless $s\bar s$ state.

\section{discussion and conclusions}

We believe is of interest to supplement lattice studies of
hadronic current correlation functions with calculations made
using chiral Lagrangian models of the type considered in this
work. We have made some progress in exhibiting results for such
correlators in Ref. [7] and in the present study. It might be of
some interest to compare our results for our temperature-dependent
RPA calculations with results obtained in the imaginary-time or
real-time formalisms, if these formalisms could be modified so
that calculations could be made in the confined phase of QCD. The
study of radial excitations in these finite-temperature theories
may be quite difficult since their study requires a model of
confinement. It is also very difficult to obtain information
concerning radial excitations, if an analytic continuation to real
time is necessary.

One interesting feature of our analysis is the use of
temperature-dependent coupling constants in the NJL model. In the
present work, we have provided some justification for the
introduction of such constants. Our work suggests that the
coupling constants of the NJL model may also be density-dependent,
since one expects that high density may play a similar role as
high temperature, leading ultimately to a weakly interacting
system at high density. We have introduced density-dependent
coupling constants in Ref. [24] where we considered the
confinement-deconfinement transition in the presence of matter.
Since the study of matter at high density is a topic of active
investigation [25-31], our suggestion of density-dependent
coupling constants may have important consequences for such
studies.

% If in two-column mode, this environment will change to single-column
% format so that long equations can be displayed. Use
% sparingly.
%\begin{widetext}
% put long equation here
%\end{widetext}

%\newpage

% Create the reference section using BibTeX:
%\bibliography{basename of .bib file}

\end{document}